\documentclass[prd,twocolumn,showpacs,amsmath,amssymb]{revtex4}
\usepackage{graphicx}

\newcommand{\rt}{\rightarrow}

\newcommand{\beq}{\begin{equation}}
\newcommand{\eeq}{\end{equation}}
\newcommand{\edq}{\end{equation}}
\newcommand{\bfg}{\begin{figure}}
\newcommand{\efg}{\end{figure}}
\newcommand{\bitm}{\begin{itemize}}
\newcommand{\eitm}{\end{itemize}}
\newcommand{\bnum}{\begin{enumerate}}
\newcommand{\enum}{\end{enumerate}}
\newcommand{\btbl}{\begin{table}}
\newcommand{\etbl}{\end{table}}
\newcommand{\btbu}{\begin{tabular}}
\newcommand{\etbu}{\end{tabular}}

\newcommand{\PPJP}{J/\psi \pi^+\pi^-}

\begin{document}

\title{\boldmath
Evidence of $\psi(3770)$ non$-D \bar D$ Decay to
$\PPJP$}
\author{
J.~Z.~Bai$^1$,        Y.~Ban$^{8}$,          J.~G.~Bian$^1$,
X.~Cai$^{1}$,          J.~F.~Chang$^1$,
H.~F.~Chen$^{14}$,    H.~S.~Chen$^1$,
J.~Chen$^{7}$,        J.~C.~Chen$^1$,     
Y.~B.~Chen$^1$
% ...
\\(BES Collaboration)\\
}

\vspace{0.2cm}

\affiliation{
\begin{minipage}{145mm}
$^1$ Institute of High Energy Physics, Beijing 100039, People's Republic of
     China\\
$^2$ China Center of Advanced Science and Technology, Beijing 100080,
     People's Republic of China\\
$^3$ Henan Normal University, Xinxiang 453002, People's Republic of China\\
$^4$ Huazhong Normal University, Wuhan 430079, People's Republic of China\\
$^5$ Hunan University, Changsha 410082, People's Republic of China\\
$^6$ Liaoning University, Shenyang 110036, People's Republic of China\\
$^7$ Nankai University, Tianjin 300071, People's Republic of China\\
$^{8}$ Peking University, Beijing 100871, People's Republic of China\\
$^{9}$ Shandong University, Jinan 250100, People's Republic of China\\
$^{10}$ Shanghai Jiaotong University, Shanghai 200030, 
        People's Republic of China\\
$^{11}$ Sichuan University, Chengdu 610064,
        People's Republic of China\\       
$^{12}$ Tsinghua University, Beijing 100084, 
        People's Republic of China\\
$^{13}$ University of Hawaii, Honolulu, Hawaii 96822\\                       
$^{14}$ University of Science and Technology of China, Hefei 230026,
        People's Republic of China\\
$^{15}$ Wuhan University, Wuhan 430072, People's Republic of China\\
$^{16}$ Zhejiang University, Hangzhou 310028, People's Republic of China\\
\vspace{0.4cm}
$^{\ast}$ Visiting professor to University of Michigan, Ann Arbor,
MI 48109 USA
\end{minipage}
}

\vspace{0.2cm}

\begin{abstract}
  Evidence of $\psi(3770)$ decays to a non-${D \bar D}$ final
  state is observed. A total of $11.8 \pm 4.8 \pm 1.3$ $\psi(3770) \rightarrow
  \PPJP$ events are obtained from a data sample
  of $27.7$ $\rm {pb^{-1}}$ taken at center-of-mass energies around 3.773
  GeV using the BES-II detector at the BEPC.  The branching fraction
  is determined to be $BF(\psi(3770) \rightarrow
  \PPJP)=(0.34\pm 0.14 \pm 0.09)\%$, corresponding to the partial
  width of $\Gamma(\psi(3770) \rightarrow \PPJP) = (80 \pm 33 \pm 23)$ keV.
\end{abstract}

\maketitle

\section{Introduction}

The $\psi(3770)$ resonance 
is believed to be a mixture of the 
$1 ^3D_1$ and $2 ^3S_1$
states of
the $c \bar c$ system~\cite{mark1}.  Since its mass is above the open
charm-pair threshold and its width is two orders of magnitude larger
than that of the $\psi(2S)$, 
it is thought to decay almost entirely to pure $D \bar
D$~\cite{Bacino}.  However, Lipkin pointed out that the $\psi(3770)$ could
decay to non$-D \bar D$ final states with a large branching
fraction~\cite{Lipkin}.  There are theoretical
calculations~\cite{Lane,Kuang1,Kuang2,Kuang3} that estimate the
partial width for $\Gamma(\psi(3770) \rightarrow \PPJP)$
based on the multipole expansion in QCD.  Recently
Kuang~\cite{Kuang3} used the Chen-Kuang potential model to obtain a
partial width for $\psi(3770) \rightarrow J/\psi \pi \pi$ in the range
from 37 to 170 keV, corresponding to 25 to 113 keV for $\psi(3770)
\rightarrow \PPJP$ from isospin symmetry.
In this paper, we report evidence for $\psi(3770) \rightarrow
\PPJP$ based on a data sample of
$27.7$ pb$^{-1}$
taken in the center-of-mass (c.m.) 
energy region from 3.738 GeV to 3.885 GeV
using the upgraded Beijing
spectrometer (BES-II) at the Beijing Electron Positron Collider
(BEPC).

\section{The BES-II detector}

The BES-II is a conventional cylindrical magnetic detector that is
described in detail in Ref.~\cite{BES-II}.  A 12-layer Vertex Chamber
(VC) surrounding the beryllium beam pipe provides input to the event
trigger, as well as coordinate information.  A forty-layer main drift
chamber (MDC) located just outside the VC yields precise measurements
of charged particle trajectories with a solid angle coverage of $85\%$
of $4\pi$; it also provides ionization energy loss ($dE/dx$)
measurements which are used for particle identification.  Momentum
resolution of $1.7\%\sqrt{1+p^2}$ ($p$ in GeV/c) and $dE/dx$
resolution of $8.5\%$ for Bhabha scattering electrons are obtained for
the data taken at $\sqrt{s}=3.773$ GeV. An array of 48 scintillation
counters surrounding the MDC measures the time of flight (TOF) of
charged particles with a resolution of about 180 ps for electrons.
Outside the TOF, a 12 radiation length, lead-gas barrel shower counter
(BSC), operating in limited streamer mode, measures the energies of
electrons and photons over $80\%$ of the total solid angle with an
energy resolution of $\sigma_E/E=0.22/\sqrt{E}$ ($E$ in GeV) and spatial
resolutions of 
$\sigma_{\phi}=7.9$ mrad and $\sigma_Z=2.3$ cm for
electrons. A solenoidal magnet outside the BSC provides a 0.4 T
magnetic field in the central tracking region of the detector. Three
double-layer muon counters instrument the magnet flux return and serve
to identify muons with momentum greater than 500 MeV/c. They cover
$68\%$ of the total solid angle.

\section{Data analysis}

\subsection{Monte Carlo simulation}
\begin{figure}[hbt]
\includegraphics*[width=9.0cm,height=9.0cm]
{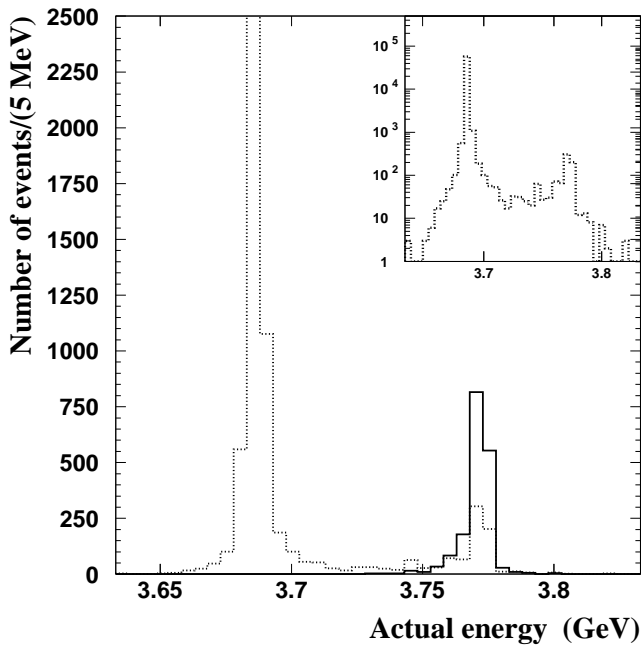}
\caption{
The numbers of $\psi(3770) \rightarrow J/\psi \pi^+\pi^-$ (solid line) and
$\psi(2S) \rightarrow J/\psi \pi^+\pi^-$ (dotted line) as a function of
the actual energy remaining after ISR, where $J/\psi$ is set to decay to
$l^+l^-$;
the events are generated with
the Monte Carlo generator at the
c.m. energies at which the data were collected
from 3.738 to 3.885 GeV. The insert on the right-top shows
the distribution of the energy 
at which $\psi(2S) \rightarrow J/\psi \pi^+\pi^-$ events are produced.
}
\label{diffxsct}
\end{figure}

To understand the main source of background in the study of the
decay $\psi(3770)\rightarrow J/\psi \pi^+\pi^-$, we developed a
Monte Carlo generator.
The Monte Carlo simulation includes the initial state
radiation (ISR) at one loop order, in which the actual center-of-mass energies after ISR
are generated according to Ref.~\cite{Kuraev}.  The $\psi(2S)$ and
$\psi(3770)$ are generated using energy dependent Breit-Wigner
functions according to Eq.(38.53) of Ref.~\cite{PDG04} in
which the ratio of $\Gamma_{el}(s)/\Gamma_{tot}(s)=\Gamma_{\psi(2S)
\rightarrow e^+e^-}/\Gamma_{tot}$ and the branching fraction of
$\psi(2S) \rightarrow J/\psi \pi^+\pi^-$ in the formula are
assumed to be constant. The beam energy spread
($\sigma_{E_{beam}}=1.37$ MeV)
is taken into account in the simulation. 
Since there is no unique description
and solution for the low energy $\pi \pi$ production
amplitude~\cite{robert_zhang_lowell},
the correction of the decay rate due to the
$\pi \pi$ production amplitude is neglected in the making of the event
generator. 
However the effect of the variations
in the correction to the decay rate on the estimated number of
$\psi(2S)\rightarrow J/\psi \pi^+\pi^-$ is considered in the final background
subtraction in subsection E.
Fig.~\ref{diffxsct} shows the distribution of $J/\psi \pi^+\pi^-$ events
with $J/\psi \rightarrow l^+l^-$ ($l=e$ or $\mu$) as a function of the
actual energy remaining after initial state photon radiation, which is determined by our Monte
Carlo generator, where the branching fraction for $\psi(3770)
\rightarrow \PPJP$ 
is set to be $0.35\%$, while the
branching fractions for 
$\psi(2S) \rightarrow \PPJP$ 
and $J/\psi \rightarrow l^+l^-$ are taken from the Particle
Data Group (PDG)~\cite{PDG04}.

There are two peaks which are around 3.686 GeV and 3.773 GeV in
Fig.~\ref{diffxsct}, where
the $J/\psi \pi^+\pi^-$ events from the $\psi(2S)$ decay are given by
the dotted histogram, while the events from the $\psi(3770)$ decay are given by
the solid histogram.
There are two components in the higher mass peak. One is from $\psi(3770)$
production, and the another is from $\psi(2S)$ production 
which is due to the tail of $\psi(2S)$ Breit-Wigner function. 
This type of $\psi(2S)$ production (called type B of $\psi(2S)$ in this
Letter) is indicated by the dotted histogram around 3.773 GeV.
The $J/\psi \pi^+\pi^-$ events in the lower mass peak are produced around
the peak of the $\psi(2S)$ Breit-Wigner function (called type A of $\psi(2S)$ in
this Letter), and are due to ISR energy return to the $\psi(2S)$ peak. 
The "type B" of $\psi(2S)$ is the main source of
background events in the experimental study of the decay 
$\psi(3770) \rightarrow J/\psi \pi^+\pi^-$. This background event has the
same topology as that as the decay of 
$\psi(3770) \rightarrow J/\psi \pi^+\pi^-$. 
To get the number of $\psi(3770)\rightarrow J/\psi \pi^+\pi^-$ signal
events, 
the number of the background
events of $\psi(2S) \rightarrow J/\psi \pi^+\pi^-$ has to be subtracted 
from the observed candidate events of $J/\psi \pi^+\pi^-$ based on analyzing the Monte
Carlo events. 

In the energy region from 3.738 to 3.885 GeV in Fig.~\ref{diffxsct},
there are 1724 $\psi(3770) \rightarrow J/\psi \pi^+\pi^-$ events and
747 $\psi(2S) \rightarrow J/\psi \pi^+\pi^-$ events. The generated events
as shown in Fig.~\ref{diffxsct} are put through the full detector
simulation based on the GEANT simulation package. The fully simulated events
are used for study of the main background. 

\subsection{Events selection}

To search for the decay of $\psi(3770) \rightarrow \PPJP$,
$J/\psi \rightarrow e^+e^-$ or $\mu^+\mu^-$, $\mu^+\mu^- \pi^+ \pi^-$ and
$e^+e^- \pi^+ \pi^-$ candidate events are selected.
These are required to have four charged tracks with
zero total charge.
Each track is required to have a good helix fit, to be consistent
with originating from the primary event vertex, and to satisfy $|\cos
\theta|<0.85$, where $\theta$ is the polar angle.

Pions and leptons must satisfy particle identification requirements.
For pions, the combined confidence level (CL), calculated for the $\pi$
hypothesis using the $dE/dx$ and TOF measurements, is required to be
greater than $0.1\%$.  In order to reduce $\gamma$ conversion
background, in which the $e^+$ and $e^-$ from a converted $\gamma$ are
misidentified as $\pi^+$ and $\pi^-$, an opening angle cut,
$\theta_{\pi^+\pi^-}>20^\circ$, is imposed.  For electron identification,
the combined confidence level, calculated for the $e$ hypothesis using
the $dE/dx$, TOF and BSC measurements, is required to be greater than
$1\%$, and the ratio
$CL_e/(CL_e + CL_{\mu} + CL_{\pi} + CL_K)$ is required
to be greater than $0.7$.
If a charged track hits the muon counter, and the
$z$ and $r \phi$ positions of the hit match with the extrapolated positions
of the reconstructed MDC track, the charged track is identified as a
muon.

The candidate events of $e^+e^-\pi^+\pi^-$ or $\mu^+\mu^-\pi^+\pi^-$
satisfying the above selection criteria are further analyzed by using two
different analysis methods to be discussed in subsection C and subsection
D.

\subsection{Analysis of the $\pi^+\pi^-$ recoil mass}

The mass recoiling against the $\pi^+\pi^-$ system is calculated using
$$M_{\rm REC}({\pi^+\pi^-}) = \sqrt{(E_{cm}-E_{\pi^+\pi^-})^2 -
|\vec P_{\pi^+\pi^-}|^2},$$
\noindent
where $E_{cm}$ is the c.m. energy,
$E_{\pi^+\pi^-}$ and $\vec P_{\pi^+\pi^-}$ are the total energy and
momentum of the $\pi^+\pi^-$ system, respectively.

\begin{figure}[hbt]
\includegraphics[width=9cm,height=9cm]
{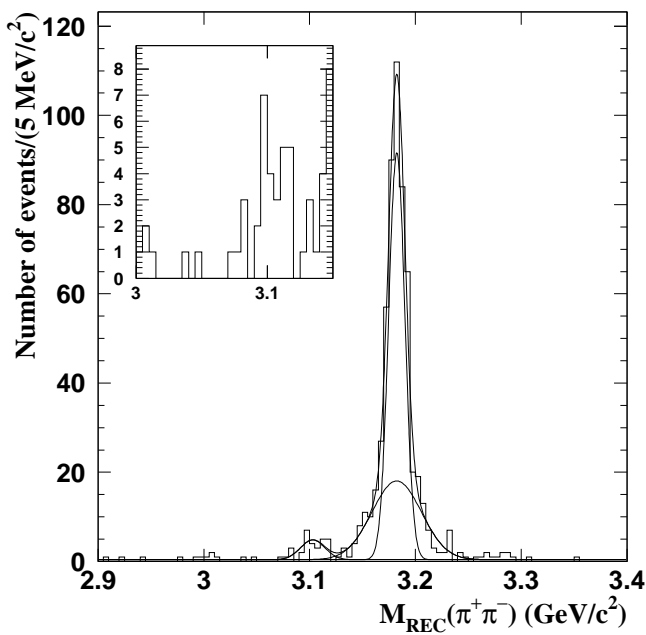}
\caption{The distribution of the masses recoiling from
the $\pi^+\pi^-$ system for $l^+l^- \pi^+\pi^-$ events;
the insert on the left-top shows the mass distribution
in a local region around 3.1 GeV; 
the curves give the best fit to the recoil mass spectrum.
see text.
}
\label{massrec1}
\end{figure}

\begin{figure}[hbt]
\includegraphics[width=9cm,height=9cm]
{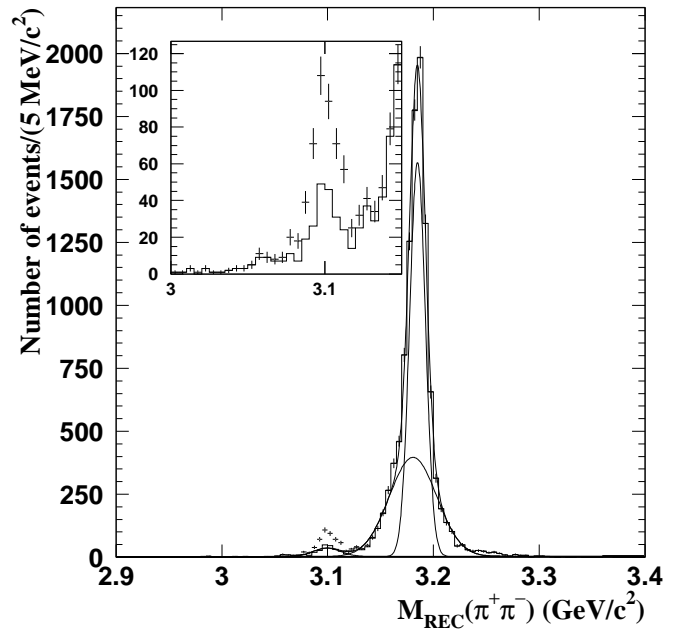}
\caption{The distribution of the masses recoiling 
from $\pi^+\pi^-$ for the
Monte Carlo events of $\psi(2S) \rightarrow  \PPJP$
and $\psi(3770) \rightarrow \PPJP$ with $J/\psi \rightarrow l^+l^-$;
these events are generated with the Monte Carlo
generator; where the 
error bars
represent the sum total of the two components,
while the histogram is for the $\psi(2S) \rightarrow J/\psi \pi^+\pi^-$
from both the "type A" and "type B" of $\psi(2S)$;
the curves give the best fits to the recoil mass spectrum from the
Monte Carlo $\psi(2S) \rightarrow J/\psi \pi^+\pi^-$ events only;
the insert on the left-top shows the mass distribution
in a local region around 3.1 GeV; 
here two components from the $\psi(3770) \rightarrow J/\psi \pi^+\pi^-$
and the "type B" of $\psi(2S) \rightarrow J/\psi \pi^+\pi^-$ are
clearly seen. 
}
\label{massrec2}
\end{figure}

Fig.~\ref{massrec1} shows the distribution of the masses recoiling against
the $\pi^+ \pi^-$ system
for candidate events with total energy
within $\pm 2.5\sigma_{E_{\pi^+\pi^-l^+l^-}}$ of the nominal
c.m. energy at which the events were obtained and with a dilepton
invariant mass within $\pm 150$ MeV of the $J/\psi$ mass,
where $\sigma_{E_{\pi^+\pi^-l^+l^-}}$ is the standard deviation of the
distribution of the energy of the $\pi^+\pi^-l^+l^-$.  Two peaks
are observed. The higher one is from the "type A" of $\psi(2S)$ events produced by
radiative return to the peak of the $\psi(2S)$,
while the small enhancement around 3.1 GeV is mostly
from $\psi(3770)$ decays, but also contains the contamination of the "type B"
of $\psi(2S)$ decays. This is confirmed by analyzing the Monte Carlo
sample generated with the Monte Carlo generator as mentioned before. 
Fig.~\ref{massrec2} shows the same distribution from analysis of the Monte
Carlo events for $\psi(2S)$ and $\psi(3770)$ production and decays to
$J/\psi \pi^+\pi^-$ with $J/\psi \rightarrow l^+l^-$ final states. The Monte
Carlo events are generated at the c.m. energies at which the data were
collected. The size of the Monte Carlo sample is 
twenty times larger than the data.
There are also two peaks; the higher one is from the "type A" of $\psi(2S)$
events;
the small enhancement
around 3.1 GeV consist of two components. 
One is from the "type B" of $\psi(2S)$ event production
and decays as shown by the solid line histogram;
the other one
is from $\psi(3770)$ production and decays
which come from the events as shown by the solid histogram in Fig.\ref{diffxsct}. 
The error bars in Fig.~\ref{massrec2} are based on 
the total number of observed events of
$\psi(3770) \rightarrow J/\psi \pi^+\pi^-$ and
$\psi(2S) \rightarrow J/\psi \pi^+\pi^-$.
The differences between the error bars and the histogram as shown around 3.1 GeV
correspond to $\psi(3770)$ production and decay to $J/\psi \pi^+\pi^-$.

Monte Carlo studies show that the distributions of the masses recoiling from
the $\pi^+\pi^-$ system for the "type A" of $\psi(2S)$ events can be
described by two Gaussian functions.
One Gaussian function is for the "type A" of $\psi(2S)$ production from the
events for which the nominal c.m. energy is set at 3.773 GeV, 
and the other one is from the events for which the nominal c.m. energies are
set off 3.773 GeV. Using triple Gaussian functions, one of which
describes the peak near the $J/\psi$ mass and two of which
represent the second and the third peaks of the events
from the "type A" of $\psi(2S)$, and a first order
polynomial to represent the background
to fit the mass distributions as shown by the solid histograms
for both the data (Fig.~\ref{massrec1}) and 
the Monte Carlo (Fig.~\ref{massrec2}) sample,
we obtain  a total of $25.5\pm 5.9$ $J/\psi \rightarrow l^+l^-$ signal events
from both the $\psi(3770) \rightarrow J/\psi \pi^+\pi^-$ and 
the "type B" of $\psi(2S) \rightarrow J/\psi \pi^+\pi^-$ events
and $220.5 \pm 26.0$ "type B" of $\psi(2S) \rightarrow J/\psi \pi^+\pi^-$ events,
respectively.
The curves give the best fits to the data and the Monte Carlo sample.
The fitted peak positions and standard deviations of the Gaussian functions
used for the fits in Fig.~\ref{massrec1} and Fig.~\ref{massrec2}
are listed in Table I.
\vspace{5mm}
\begin{table}
\centering
\vspace{0.2cm}
\caption{Summary of the fitted results of the data and the Monte Carlo
sample in Fig.~\ref{massrec1} and Fig.~\ref{massrec2}.}
\begin{tabular}{cccc}
 \hline \hline
Figure & Peak & Mass [MeV]  &  $\sigma_{M}$ [MeV] \\ \hline
\hline
       & Peak1 & $3097.8 \pm 3.0$ & $9.9 \pm 2.4$ \\
\ref{massrec1}   & Peak2  & $3182.4 \pm 1.8$ & $23.3 \pm 2.8$ \\
       & Peak3 & $3182.5 \pm 0.6$ & $7.6\pm0.7$ \\
\hline
       & Peak1 & $3099.4 \pm 1.3$ & $13.1 \pm 2.9$ \\
\ref{massrec2}   & Peak2  & $3180.8 \pm 0.4$ & $22.4 \pm 0.7$ \\
       & Peak3   & $3185.0 \pm 0.2$ & $7.4\pm0.2$ \\
\hline \hline
\end{tabular}
\end{table}

\subsection{Kinematic fit}

\begin{figure}
\includegraphics[width=9.0cm,height=9cm]
{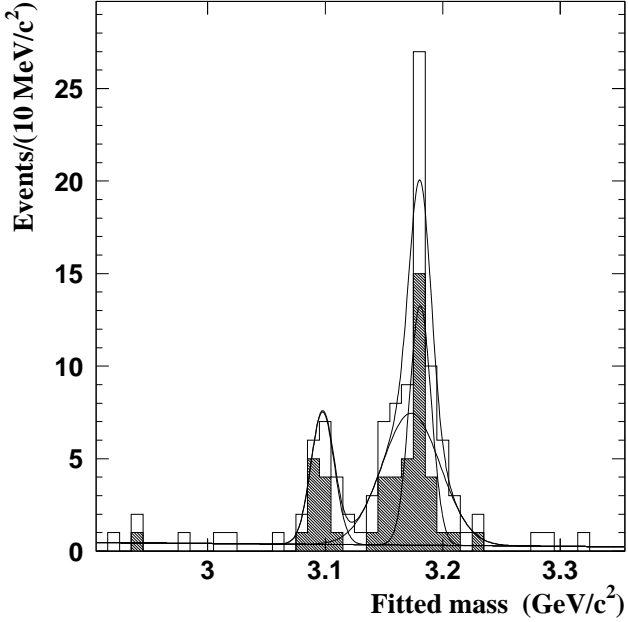}
\caption{The distribution of the fitted dilepton masses for the
events of $l^+l^-\pi^+\pi^-$ from the data;
the hatched histogram is for $\mu^+\mu^-\pi^+\pi^-$,
while the open one is for $e^+e^-\pi^+\pi^-$; the curves give the best fit
to the data.
}
\label{fitmassplot1}
\end{figure}

In order to reduce background and improve momentum resolution,
candidate events are subjected to four-constraint kinematic fits to
either the $e^+e^- \rightarrow \mu^+\mu^- \pi^+ \pi^-$ or the $e^+e^-
\rightarrow e^+e^- \pi^+ \pi^-$ hypothesis.  Events with a confidence
level greater than 1\% are accepted.  Fig.~\ref{fitmassplot1} shows the dilepton
masses determined from the fitted lepton momenta of the accepted
events.  There are clearly two peaks. The lower mass peak is mostly
due to $\psi(3770) \rightarrow J/\psi \pi^+\pi^-$, while the higher
one is due to the "type A" of $\psi(2S) \rightarrow J/\psi \pi^+\pi^-$. 
Since the higher mass peak is produced by the radiative return to the $\psi(2S)$ peak,
its energy will be approximately 3.686 GeV, while the c.m. energy is
set to the nominal energy in the kinematic fitting.  Therefore,
the dilepton masses calculated based on the fitted lepton momenta from
$\psi(2S) \rightarrow \PPJP$, $J/\psi \rt l^+ l^-$ are shifted
upward to about 3.18 GeV.

A maximum likelihood fit
to the mass distribution in Fig.~\ref{fitmassplot1}, 
using three Gaussian functions to describe
the mass distribution of the $l^+l^- \pi^+\pi^+$ combinations
and a first order polynomial to represent the broad background
as used to fit the $\pi^+\pi^-$ recoil mass distributions in
Fig.~\ref{massrec1} and Fig.~\ref{massrec2},
yields a $J/\psi$ mass value of $3097.8\pm 3.0$ MeV
and  a signal of 
$17.8\pm 4.8$ $J/\psi \rightarrow l^+l^-$
events. The curves give the best fit to the data.

\begin{figure}
\includegraphics[width=9.0cm,height=9cm]
{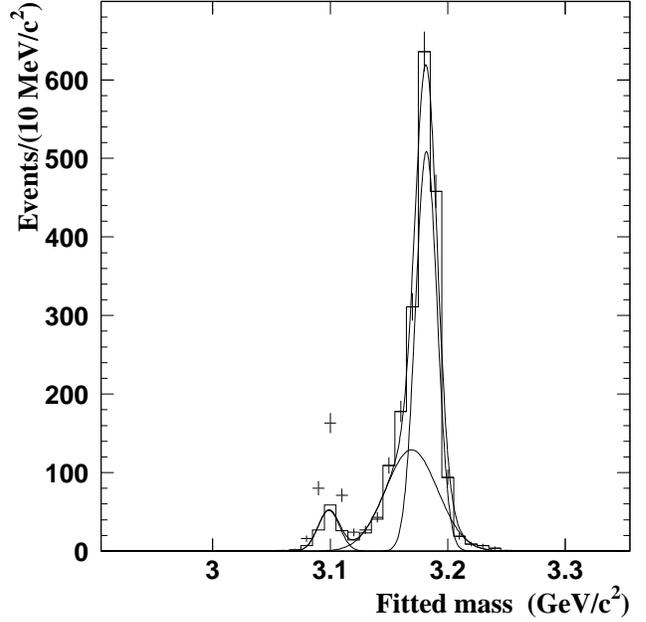}
\caption{
The distribution of the fitted dilepton masses for the
events of $l^+l^-\pi^+\pi^-$ from the 
Monte Carlo sample of $\psi(3770) \rightarrow J/\psi \pi^+\pi^-$
and $\psi(2S) \rightarrow J/\psi \pi^+\pi^-$ which are generated with the
Monte Carlo generator (see section A); 
the histogram is for $\psi(2S)\rightarrow J/\psi \pi^+\pi^-$, 
while the 
error bars are the sum of
$\psi(3770) \rightarrow J/\psi \pi^+ \pi^-$ and 
$\psi(2S) \rightarrow J/\psi \pi^+\pi^-$, 
where the $J/\psi$ is set to decay to $l^+l^-$.
}
\label{fitmassplot2}
\end{figure}

As discussed in Section C,
there is a contribution from the "type B" of
$\psi(2S) \rightarrow \PPJP$
that can pass the event selection criteria and can lead 
to an accumulation of
the recoil masses of the $\pi^+\pi^-$
and/or the fitted dilepton masses
around 3.097 GeV.  This is the main source of background to
$\psi(3770) \rightarrow \PPJP$.
Fig.~\ref{fitmassplot2} shows the distribution of the
fitted dilepton masses of the Monte Carlo events of
$\psi(3770) \rightarrow J/\psi \pi^+\pi^-$ and
$\psi(2S) \rightarrow J/\psi \pi^+\pi^-$ with $J/\psi \rightarrow l^+l^-$
as shown in Fig.~\ref{diffxsct}.
Here the histogram shows the dilepton mass distribution for
$\psi(2S) \rightarrow J/\psi \pi^+\pi^-$ only.  The higher mass peak is due
to the "type A" of $\psi(2S) \rightarrow J/\psi \pi^+\pi^-$ events,
and the lower one is from the "type B" of
$\psi(2S) \rightarrow J/\psi \pi^+\pi^-$ events.
The error bars show the 
sum total of the observed events of
$\psi(2S) \rightarrow J/\psi \pi^+\pi^-$ and
$\psi(3770) \rightarrow J/\psi \pi^+\pi^-$.
The differences between the error bars and the histogram correspond
to the 
observed events of
$\psi(3770) \rightarrow J/\psi \pi^+\pi^-$.
Fitting the mass distribution for the $\psi(2S)$ events
only (histogram) with the same triple Gaussian functions
as mentioned before yields
$119 \pm 12.1$ $J/\psi$ events from the "type B" of
$\psi(2S) \rightarrow J/\psi \pi^+\pi^-$ decays.

\begin{figure}
\includegraphics[width=9.0cm,height=9.0cm]
{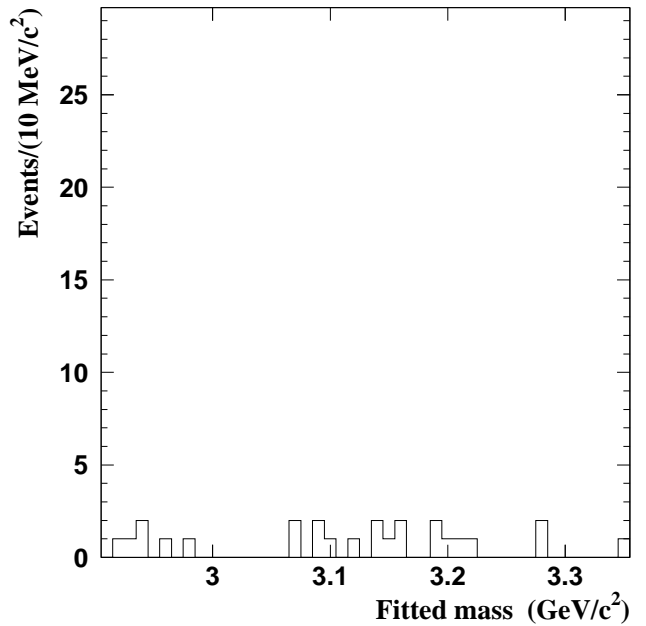}
\label{dlpmass_4030gev}
\caption{The distributions of the dilepton masses of the events
of $l^+l^-\pi^+\pi^-$ from the data taken at 4.03 GeV with the BES-I at the BEPC.
}
\label{mfit_4030mev}
\end{figure}

\begin{figure}
\includegraphics[width=9.0cm,height=9cm]
{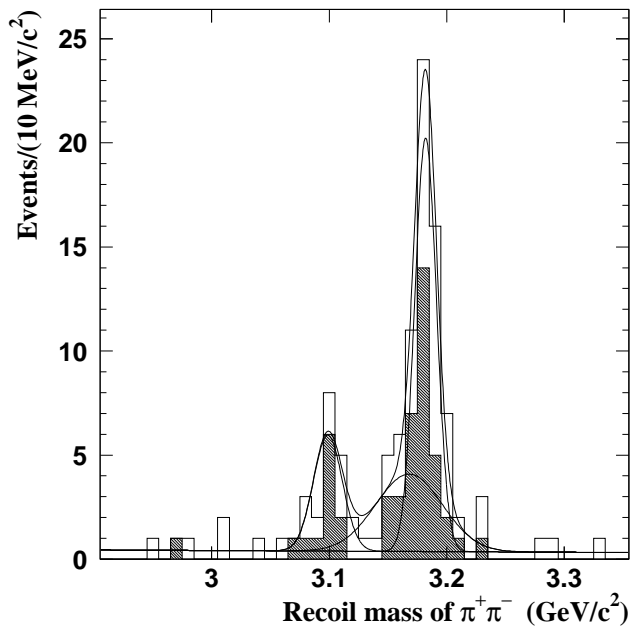}
\caption{
The distribution of the masses recoiling against the
$\pi^+\pi^-$ system calculated using the measured momenta for events that
pass the kinematic fit requirement, where the hatched histogram
is for events of $\mu^+\mu^-\pi^+\pi^-$ and the open one is for
$e^+e^-\pi^+\pi^-$.
}
\label{fitmassplot3}
\end{figure}

\subsection{Other background and background subtraction}

\subsubsection{Other background}

Some physics processes, such as two-photon events, $e^+e^- \rightarrow
e^+e^-\mu^+\mu^-$ (where the slow muons are misidentified as pions) and
$e^+e^- \rightarrow e^+e^-\pi^+\pi^-$, 
$e^+e^- \rightarrow \tau^+\tau^-$
and
$e^+e^- \rightarrow D \bar D$ could be sources of background.

To check if there are some background contaminations in the observed
$J/\psi \pi^+\pi^-$ events due to the possible sources of background,
we generated $1\times 10^5$ two-photon Monte Carlo events
(which is about 4 times larger than the data), 
$6\times 10^5$ $e^+e^- \rightarrow hadrons$ Monte Carlo events
(which is about 1.6 times larger than the data),
and $2.3\times 10^6$ $e^+e^- \rightarrow D \bar D$ Monte Carlo events
(which is about 13 times larger than the data),
where the $D$ and $\bar D$ mesons are set 
to decay to all possible final states
according to the decay modes and branching fractions
quoted from PDG~\cite{PDG04}. These Monte Carlo events are fully simulated 
with the GEANT-based simulation package. 
None of the simulated possible background
events were misidentified as $J/\psi \pi^+\pi^-$ events.

The candidate $\PPJP$ events could also
be produced in the continuum process, 
such as $e^+ e^- \rightarrow l^+l^-\pi^+\pi^-$ and
$e^+e^- \rightarrow \tau^+\tau^-$, 
and satisfy the selection criteria.
From analyzing a sample of 6.6 pb$^{-1}$ taken
at 3.65 GeV with the BES-II detector,
a sample of 5.1 pb$^{-1}$ taken in the energy region from 3.544
to 3.600 GeV and
a sample of  22.3 $\rm pb^{-1}$ taken at 4.03 GeV
with the BES-I detector,
no significant $\PPJP$, $J/\psi \rightarrow l^+ l^-$ events are observed.
Fig.~\ref{mfit_4030mev}
shows the distribution of the fitted dilepton masses
of the events of $l^+l^-\pi^+\pi^-$ which satisfy the selection criteria;
these events are from the data taken with the BES-I detector at 4.03 GeV.
The distribution of the fitted dilepton masses is flat, which is consistent
with the background distribution.
Hence the continuum background is negligible.

\subsubsection{Number of background}

After normalizing to the total
luminosity of the data set, we estimate that there are
$11.0 \pm 1.3 \pm 2.4$ background events from the
"type B" of $\psi(2S) \rightarrow J/\psi \pi^+\pi^-$
in the $25.5\pm 5.9$ $J/\psi \rightarrow l^+l^-$ signal events obtained by
fitting to the $\pi^+\pi^-$ recoil mass distribution of Fig.~\ref{massrec1}
and
$6.0 \pm 0.5 \pm 1.3$ background events from the
"type B" of $\psi(2S) \rightarrow J/\psi \pi^+\pi^-$
in the $17.8\pm 4.8$ $J/\psi \rightarrow l^+l^-$ signal events obtained by
fitting to the fitted dilepton mass distribution of
Fig.~\ref{fitmassplot1}, where the first errors are statistical and 
the second are systematic. The later arise from the uncertainties ($\pm 1.1$) and ($\pm 0.6$)
in the $\psi(2S)$ resonance parameters and
the uncertainties ($\pm 2.2$) and ($\pm 1.2$) coming from the ambiguities of
the knowledge of the low energy $\pi \pi$ production amplitude in 
$\psi(2S) \rightarrow J/\psi\pi^+\pi^-$.
The two terms correspond to the uncertainty on the production amplitute  
predicted by different theoretical models~\cite{robert_zhang_lowell}
for analyzing the $\pi^+\pi^-$ recoil mass spectrum and the fitted dilepton
mass spectrum, respectively.
The theoretical models are based on the PCAC and current algebra or
chiral pertubative theory predictions in which the various $\pi^+\pi^-$
rescatering corrections are taken into account to get
better unitarity behavior at higher energy.

\subsection{Number of signal events 
$\psi(3770) \rightarrow J/\psi \pi^+\pi^-$}

The probability that the 17.8 events observed
are due to a fluctuation of the $6.0 \pm 0.5 \pm 1.3$ events is
$1.1\times 10^{-3}$.
After subtracting the numbers of the background events, 
$14.5\pm 6.4 \pm 2.4$ and
$11.8 \pm 4.8 \pm 1.3$ signal events of
$\psi(3770) \rightarrow J/\psi \pi^+\pi^-$ are retained
from analyzing the recoil masses of $\pi^+\pi^-$ (see Fig. 2) 
and the fitted dilepton masses (see Fig.~\ref{fitmassplot1}) of the events
$l^+l^-\pi^+\pi^-$, respectively.

In this analysis, the possible interference between the $\psi(2S)$
background and the $\psi(3770)$ signal is neglected since the
decay wave functions of $\psi(3770)$ and $\psi(2S)$ are orthogonal
~\cite{Kuang2}. The BES detector is symmetric enough in the spatial
direction and there is no bias for the event selections about the
momentum direction of the particles. 
Therefore the interference terms cancel
after integrating over the pion momenta.

To test whether there is any bias in the kinematic fit, we examine the
$\pi^+\pi^-$ recoil mass distribution for the events passing the kinematic
fit requirements.
Fig.~\ref{fitmassplot3} shows the recoil mass distribution,
where the recoil masses are calculated as mentioned in Section C,
but the events are not required to satisfy the total energy cut and dilepton
invariant mass cut.
There are also two peaks, similar to those in Fig.~\ref{fitmassplot1}, observed clearly.
Fitting to the mass spectrum with
the same functions as described above
yields a $J/\psi$ mass value of $3100.1\pm 3.8$ MeV
and a signal of 
$17.2\pm 5.0$
events,
consistent with the $17.8 \pm 4.8$ 
signal events 
obtained by fitting to the dilepton mass distribution in
Fig.~\ref{fitmassplot1}. 

Table II summarizes the fitted peak positions and standard deviations
of the Gaussian functions used for the fits in 
Fig.~\ref{fitmassplot1}, Fig.~\ref{fitmassplot2} and Fig.~\ref{fitmassplot3}.

\vspace{5mm}
\begin{table}
\centering
\vspace{0.2cm}
\caption{Summary of the fitted results of the data and Monte Carlo sample in
Fig.~\ref{fitmassplot1}, Fig.~\ref{fitmassplot2} and Fig.~\ref{fitmassplot3}.}
\begin{tabular}{cccc}
 \hline \hline
Figure & Peak & Mass [MeV]  &  $\sigma_{M}$ [MeV] \\ \hline
\hline
       & Peak1 & $3097.8 \pm 3.0$ & $9.9 \pm 2.4$ \\
Fig.~\ref{fitmassplot1}
       & Peak2  & $3173.1 \pm 5.5$ & $24.8 \pm 5.7$ \\
       & Peak3  & $3180.9 \pm 2.4$ & $9.3$ (fixed)  \\
\hline
       & Peak1 & $3098.8 \pm 0.7$ & $9.1 \pm 0.5$  \\
Fig.~\ref{fitmassplot2}
   & Peak2     & $3169.2 \pm 0.7$ & $22.2 \pm 0.5$  \\
       & Peak3 & $3181.8 \pm 0.2$ & $9.3$ (fixed) \\ \hline
       & Peak1 & $3100.1 \pm 3.8$ & $13.2 \pm 3.5$  \\
Fig.~\ref{fitmassplot3}
       & Peak2   & $3177.1 \pm 3.9$ & 19.6 (fixed)  \\
       & Peak3   & 3185.4 (fixed) & 7.0 (fixed) \\
\hline \hline
\end{tabular}
\end{table}

\subsection{The number of $\psi(3770)$ produced}

The total number of $\psi(3770)$ events is obtained from our measured
luminosities at each c.m. energy and from calculated cross sections
for $\psi(3770)$ production at these energies.  The Born level
cross section at energy $E$ is given by
$$\sigma_{\psi(3770)}^{B}(E) = \frac{12 \pi \Gamma_{ee}\Gamma_{\rm tot}(E)}
{{(E^2-M^2)^2 + M^2\Gamma^2_{\rm tot}(E)}},$$
where the $\psi(3770)$ resonance parameters, $\Gamma_{ee}$ and
$M$, are taken from the
PDG~\cite{PDG04} and 
$\Gamma_{\rm tot}(E)$ is chosen to be energy dependent and normalized to
the total width $\Gamma_{\rm tot}$ 
at the peak of the resonance~\cite{PDG04}\cite{MARKII}\cite{bes2_ddbxsct}.
In order to obtain the observed cross section,
it is necessary to correct for ISR.
The observed $\psi(3770)$ cross section,
$\sigma^{obs}_{\psi(3770)}(s_{nom})$, is reduced by a factor $ g(s_{nom}) =
{\sigma^{obs}_{\psi(3770)}(s_{nom})}/
{\sigma^{B}_{\psi(3770)}(s_{nom})}$, where $s_{nom}$ is the
c.m. energy squared and ${\sigma^{B}_{\psi(3770)}(s_{nom})}$
is the Born cross section. The ISR correction for
$\psi(3770)$ production is calculated using a Breit-Wigner function
and the radiative photon energy spectrum~\cite{Kuraev}\cite{bes2_ddbxsct}.
With the calculated cross sections for $\psi(3770)$ production at each
energy point around 3.773 GeV and the corresponding luminosities,
the total number of $\psi(3770)$ events
in the data sample is determined to be
$ N_{\psi(3770)}^{prod} =  (1.85 \pm 0.37) \times 10^5$,
where the error is mainly due to the uncertainty in the
observed cross section for $\psi(3770)$ production.

\section{Result}

\subsection{Monte Carlo efficiency}

The efficiencies for reconstruction of the events of
$\psi(3770)\rightarrow \PPJP$ with $J/\psi \rightarrow e^+e^-$ and
$J/\psi \rightarrow \mu^+\mu^-$ are estimated by Monte Carlo simulation.
Monte Carlo study shows that the efficiencies are
$\epsilon_{\psi(3770)\rightarrow \PPJP,J/\psi \rightarrow 
e^+e^-}=0.146\pm0.003$ 
and
$\epsilon_{\psi(3770)\rightarrow \PPJP,J/\psi \rightarrow
\mu^+\mu^-}=0.174\pm0.003$,
where the errors are statistical.
These give the averaged efficiency for detection of 
$J/\psi \rightarrow e^+e^-$ and $J/\psi \rightarrow \mu^+\mu^-$
events to be $\epsilon_{\psi(3770) \rightarrow  \PPJP,
J/\psi \rightarrow l^+l^-}=0.160 \pm 0.002$.

\subsection{Branching fraction and partial width}

Using these numbers and the known branching fractions for $J/\psi
\rightarrow e^+e^-$ and  $\mu^+\mu^-$~\cite{PDG04}, the branching
fraction
for the non$-D \bar D$ decay
$\psi(3770) \rightarrow \PPJP$ is measured to be
$$BF(\psi(3770) \rightarrow \PPJP) =
 (0.338 \pm 0.143 \pm 0.086) \%, $$
where the first error is statistical and the second systematic
arising from 
the uncertainties in the total
number of $\psi(3770)$ produced ($20 \%$ ),
tracking efficiency ($2.0\%$ per track), 
particle identification $(2.2\%)$,
background shape ($6 \%$),
$\psi(2S)\rightarrow J/\psi \pi^+\pi^-$ background subtraction
($12 \%$) and
the averaged branching fraction for
$J/\psi \rightarrow l^+l^-$ ($1.2\%$).
Adding these uncertainties in quadrature yields the total systematic
error of $25.5\%$.

Using $\Gamma_{\rm tot}$ from the PDG~\cite{PDG04},
this branching fraction corresponds to a partial width of
$$\Gamma(\psi(3770) \rightarrow \PPJP) =
(80 \pm 33 \pm 23)~~{\rm {keV}}, $$
where the first error is statistical and the second systematic.
The systematic uncertainty in the measured partial width arises from
the systematic uncertainty in the measured branching fraction ($25.5\%$)
and the uncertainty in the total width of $\psi(3770)$
($11.5\%$)~\cite{PDG04}.

As a consistency check, we can use the number of
signal events $\psi(3770) \rightarrow J/\psi \pi^+\pi^-$
obtained from analyzing the recoil masses of the $\pi^+\pi^-$ from the
events $l^+l^-\pi^+\pi^-$ to calculate the branching fraction.
The detection efficiency is
$\epsilon_{\psi(3770) \rightarrow  \PPJP,
J/\psi \rightarrow l^+l^-}=0.194$.
These numbers yield a
branching fraction
$BF(\psi(3770) \rightarrow \PPJP) = (0.342 \pm 0.142 \pm 0.097)\%$, which is
in good agreement with the value
obtained above, indicating that the kinematic fit result is
reliable.

\section{Summary}

In summary, the branching fraction for 
$\psi(3770) \rightarrow \PPJP$
has been measured.
From a total of $(1.85 \pm 0.37) \times 10^5$ $\psi(3770)$ events,
$11.8 \pm 4.8\pm 1.3$ non$-D \bar D$ decays of
$\psi(3770) \rightarrow \PPJP$ events
are observed, leading to a branching fraction of
$BF(\psi(3770) \rightarrow \PPJP)=
(0.34 \pm 0.14 \pm 0.09)\%$, and a partial width
$\Gamma(\psi(3770) \rightarrow \PPJP) =
(80 \pm 33 \pm 23)~{\rm {keV}}$. 

\vspace{1.5cm}
\begin{center}
{\small {\bf ACKNOWLEDGEMENTS}}
\end{center}
\par
\vspace{0.4cm}

We would like to thank the BEPC staff for their strong efforts
and the members of the IHEP computing center for their helpful
assistance. We acknowledge Professor Yu-Ping Kuang and 
Professor Kuang-Ta Chao for many helpful discussions on the
non$-D \bar D$ decay of $\psi(3770)$. 
This work is supported in part by the National Natural Science Foundation
of China under contracts
Nos. 19991480, 10225524, 10225525, the Chinese Academy
of Sciences under contract No. KJ 95T-03, the 100 Talents Program of CAS
under Contract Nos. U-24, U-25, and the Knowledge Innovation Project of
CAS under Contract Nos. U-602, U-34(IHEP); by the National Natural Science
Foundation of China under Contract No. 10175060(USTC); and by the Department
of Energy under Contract No. DE-FG03-94ER40833 (U Hawaii).


\end{document}